# Presentation an Approach for Optimization of Semantic Web Language Based on the Document Structure

Farzad Parseh[1], Davood Karimzadgan Moghaddam[2], Mir Mohsen Pedram[3],
Rohollah Esmaeli Manesh[4], Mohammad(behdad) Jamshidi[5]
[1,2]Department of IT and Communication, Payam Noor University, Tehran, Iran.
[3]Engineering Department, Faculty of Engineering, Tarbiat Moallem University, Karaj/Tehran, Iran.
[4]Young Researchers Club, Gilan Ghab Branch, Islamic Azad University, GilanGharb, kermanshah, Iran.
[5]Young Researchers Club, Kermanshah Branch, Islamic Azad University, Kermanshah, Iran.

**ABSTRACT**

Pattern tree are based on integrated rules which are equal to a combination of some points connected to each other in a hierarchical structure, called *Enquiry Hierarchical (EH)*. The main operation in pattern enquiry seeking is to locate the steps that match the given EH in the dataset. A point of algorithms has offered for EH matching; but the majority of this algorithms seeks all of the enquiry steps to access all EHs in the dataset. A few algorithms such as seek only steps that satisfy end points of EH. All of above algorithms are trying to locate a way just for investigating direct testing of steps and to locate the answer of enquiry, directly via these points. In this paper, we describe a novel algorithm to locate the answer of enquiry without access to real point of the dataset blindly. In this algorithm, first, the enquiry will be executed on enquiry schema and this leads to a schema. Using this plan, it will be clear how to seek end steps and how to achieve enquiry dataset, before seeking of the dataset steps. Therefore, none of dataset steps will be seek blindly.

**Keywords**: *Pattern, Branch Links, Query Indicators and Evaluation.*

## 1. INTRODUCTION

Enquiry seeking is an essential part of any point base. Both *XQuery* and *XEnquiry*, the two most popular enquiry rules in pattern domain, are based on *integrated rules*. A integrated rule specifies patterns of predicates selection on multiple points that has a tree schema named *Enquiry Hierarchical (EH)*. Consequently, in order to seek pattern trees, all occurrences of EH in the pattern dataset should be found. This is an expensive task when huge pattern dataset are involved. Consider the following enquiry: *Q1: //book[.//title//xml]//author//jane;* The schema of an pattern enquiry could be shown in a EH, for example the EH of enquiry Q1 is presented in Figure 1.

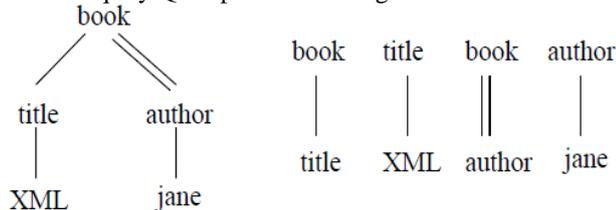

Figure 1. EH Pattern

The aim of all pattern enquiry seeking algorithms is to locate all EH instances in the pattern dataset. A point of algorithms are proposed to answer trees link. We classify this algorithms into three parts:

*part A*: Algorithms in this part are based on a famous algorithm named *path Link* [1]. In path Link, enquiry is decomposed into some binary link operations. Thus, a huge volume of intermediate dataset are produced in this algorithms.

*part B*: *Holistic branch link* algorithms[2] does not decompose the enquiry into its binary *Parent-Child (P-C)* or *Ancestor-Descendant (A-D)* relationships but they need to seek all of the enquiry steps in the dataset.

*part C*: It is better to seek only steps that satisfy ends nodes of EH. [12] is such an a algorithm encoding. (see figure 2)

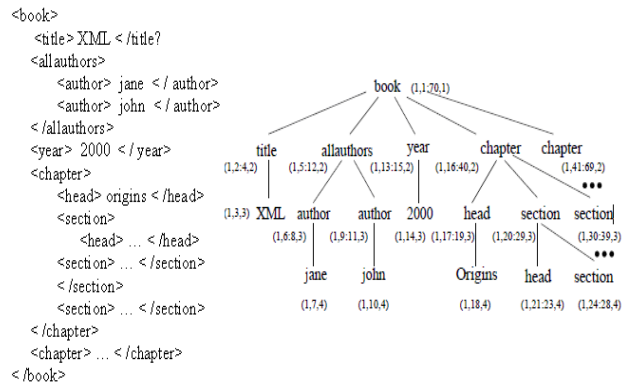

Figure 2. Schema Encoding

Three parts above called Shcema Encoding. containment link Containment link algorithms use an index named *Name indicators* to quick access to points which have same tag name. for example to answer *Q*uery, this index makes it possible to access to all steps in the dataset; but all of algorithms above, do not consider the place of points. They are trying to locate a way just for investigating direct testing of steps and to get the answer of enquiry, directly via these tests where as many of these test do not produce any part of the enquiry answer.

On the other hand, there are some *query indicators* link Strong *PointGuide, Fabric Index, ToXin, APEX, Index1, A(k) Index,* and *F&B* which are indexing the query of dataset's steps to facilitate access to steps required in pattern enquiry seeking Algorithms[3,6,7,10,13,14].
These query indicators are other kinds of enquiry seeking algorithms which are against the *A, B* and *C* part algorithms. query indicators usually have two parts:
• *Path Guide (PG)* that summarizes dataset schema and describes relation between points. (see figure 3)
• *Records* that keeps real point of the dataset based on Path Guide.

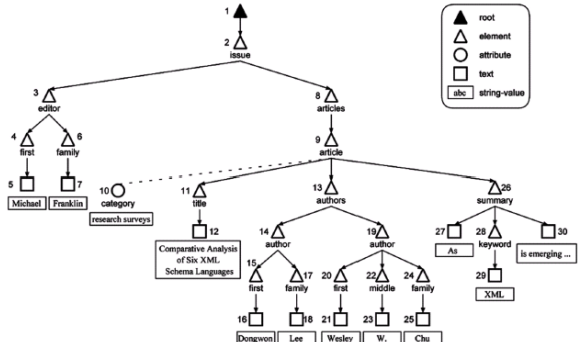

Figure 3. PG Structure

All of algorithms in this part behave as follows: At first, path relationship (A-D or P-C) between enquiry points are tested with Path Guide. As Data, Records of steps that match with enquiry is returned. For example in Match Seeking of enquiry, step point *8* matches with enquiry. Therefore, all of its Records will be returned as Dataset. This algorithm is considerable because it apply enquiry on a small set named PG and to execute the enquiry it doesn't need to access to real point of the dataset; But always trees are not such simple. For example to answer the trees such as *a//b[c]* or *a[.//b]/c* they need to access real point of the dataset. Therefore, this algorithm has not enough performance.

None of the Shcema Encoding algorithms uses full potential of query indicators or path summaries, while these have great potential to guide us to sigh seeking.

In this paper, we propose a compound algorithm that uses schema summary as enquiry schema. In this algorithm, enquiry will be executed on schema summary that has very small size in test is on with the dataset. For this purpose, there is no need to access to real point of the dataset. Data of this execution is generation of a schema called *DataTable (DT)*. DT shows end steps of the enquiry and the way of their seeking in the dataset. This save us from direct and blind seeking in the dataset.

## 2. OVERVIEW OF OUR ALGORITHM

Our algorithm is similar to both *Schema Encoding* and *Query Indicators*. In this algorithm, we apply the enquiry on Path Guide of the dataset. PG is similar to schema of a dataset and has not close relation with size of the dataset. Its size and schema are usually stable or with a few variation. (see figure 4)

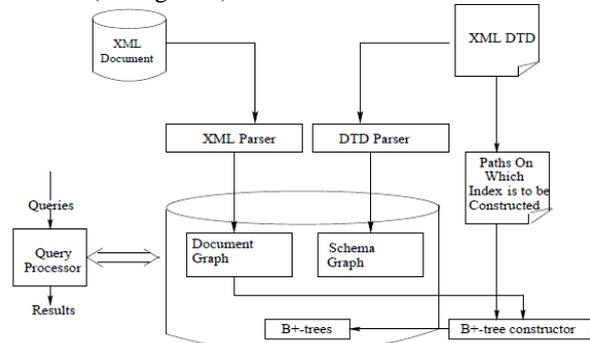

Figure 4. Data Model

*Step1*: link Query Index algorithms, first, enquiry is applied on PG; but here the enquiry is not executed in its complicated form. It will be split in several single-branch trees that will be easily answered in all algorithms of query indicators[3] [6] [7] [9] [10] [13] [14].

*Step2*: all single-branch trees execute on PG separately. A schema that called Data Table is build from execution Data of single-branch trees. DT as seek schema shows the end steps that are to seek and the way of seeking them in the dataset.

*Step3*: The dataset is numbered base on Hierarchical encoding.

*Definition*: In Hierarchical labeling algorithm if step $U$ is the $n^{th}$ child of step $V$, the Hierarchical code of step $U$ is the Hierarchical code of step $V$ as its prefix continue with $n$, Hierarchical $(U)$= Hierarchical $(V)$+'.'+'$n$'. For example suppose that Hierarchical $(V)$=<1.3> and step $U_1$ is the $7^{th}$ child of step $V$, then Hierarchical $(U_1)$=<1.3.7>

Based on Hierarchical numbers, all steps corresponding to each step of PG are sorted in Records. Third step is similar to Containment links algorithms. Based on DT, end steps of enquiry that are placed in Records will be tested and final Data will be generated.

### 2.1 ENQUIRY SPLITTING AND EXECUTION OF SINGLE-BRANCH TREES ON PATH SUMMARY

Trees are usually complicated and several-branch. Before splitting a enquiry in several single-branch trees, we should be familiar with *link point* concept.

*Definition-JP:* Link Point is a step in EH which links more than one branch to each other.

*Example:* suppose $A$ and $B$ are two branches of a enquiry that have traversed query from enquiry root $a1/a2/.../aj/ax1/.../an$ and $a1/a2/.../aj/ax2/.../am$ and $ax1 \neq ax2$ then $J$ is link point of two branches with $a1/a2/.../aj$ as its query. We do not mean parent-child relation by / between enquiry points and it can be interpreted as /, //, *.
To answer the several-branch trees, we need to locate link points of branches that called JP. Complexity of several-branch trees is because of JPs. We can easily locate place of these steps on PG; but we cannot definitely answer to this kind of enquiry without access to the dataset. Enquiry condition is as follow: a JP in a dataset is part of answer if it has all of enquiry branches under itself, in other words, several enquiry branches in the dataset can be part of answer if they are link in same JP. This JP cannot be found just with access to PG and without testing of branches in the dataset; because it is possible that one JP in the dataset has not one of enquiry branches under itself.
*Example:* *A* is link point of two branches, *A//B* and *A//C//D*. *A* steps in dataset are part of answer if have both of *A//B* and *A//C//D* branches.
*Splitting Enquiry:* suppose *Q* is a several-branch enquiry with *n* JPs and m branches (ends nodes). *Q* split in single-branch trees $SQ_1, ...., SQ_m$ so that each $SQ_i$ is a branch from root to end of one of branches and every two of $SQ_i$ and $SQ_j$ have same prefix from root to one of the JPs. Total point of these different JPs is *n*.

Here our goal is description of algorithm functionality. For this reason, we explain our algorithm on simple enquiry and then we show how DT can answer to complicated trees.
*The procedure*: As mentioned above, at first, we must split enquiry. Enquiry split into single-branch trees. Then each single-branch enquiry will be executed on PG separately. Fortunately, in most of query index algorithms single-branch trees can be answered easily with PG and without access to the dataset point. Data of this execution will be a list of steps in PG for each single-branch enquiry. Query of these steps will be absolute (from root to step in PG).
*Example:* suppose we want to execute enquiry on PG. at first, enquiry split into two single-branch trees: *A//B* and *A//C//D*. we only need to keep and access to ends nodes of enquiry for each branch because the dataset labeled with Hierarchal numbers and lower steps have some information about upper steps (query traversed from root) in themselves.

## 2.2 GENERATION OF DT

*Primary Definition:* DT is a table with three columns. First two columns are end steps of two enquiry branches in PG and its third column is step of JP between two these branches. End steps in DT have absolute query. Therefore, each record of this table shows an operation called *Matching Seek*.
*Definition:* Matching Seek is seek of testing two or more steps in the dataset to achieve part of answer.

*The procedure:* after splitting enquiry into several single-branch trees and gaining corresponding steps to ends nodes of single-branch trees in PG, now we have to achieve JP of these steps. In Hierarchal encoding manner, each end indicate a branch. Data of single-branch trees execution on PG is a list of steps for each single-branch enquiry. The Query of these steps are absolute (i.e, query of each step is completely specified from root to step). Now, to achieve JP of these steps, we select a step from each list and test their absolute querys with each other. If querys of selected steps were same from root to step of enquiry JP, we add those two steps and step of JP to DT.

```
For each jp1 in an, jp2 in bn do
If an.prefix(jp1)=bn.prefix(jp2) then
DT.addREC(an, bn, jp1.Level)
```

## FINAL DATA

Final Data is constructed based on the DataTable. Each record in the DataTable guides enquiry seek to produce a part of the final Data. Therefore, final Data is the union of partial Datas produced for each record of DataTable.
*The procedure:* Consider a given record in a DataTable and its fields. Two first fields are two steps in a Schema Summary. As mentioned in introduction, each step in Schema Summary has an ordered list of related steps' *Hierarchical* point in the pattern dataset that called Records. Points of these two lists should be tested with each other to produce part of the final Data. This seek is called Matching Seek. The matching seek starts with testing current step labels of lists (first ones at the beginning). If testing steps have same prefix up to JP step (third field), those are part of Data.
*Example:* Consider record *B1, D1* of the previous DataTable (the JP value of the record is assumed *2*). Suppose their related step labels form the below lists:
Step of *W* is assumed *ε*.
The three bolded Lines give us steps that have same prefix up to JP step and are part of Matching Seek Datas. For steps such as *1/2/2/1* which have not successful matching seek, we should jump to next first step that is just greater in this step (look at *jump(L)*). For example in step *2*, if step *1/2/2/1* is current step, then next step will be *1/3/3/1*.

```
For each a in L1 ,b in L2 do
a.prefix(L)=b.prefix(L) then
(a,b) add to output
node1.prefix(L)>node2.prefix(L) then
node2=L2.Jump(L)
node1=L1.Jump(L)
```

## 3. DT AND COMPLICATED TREES

In previous sections, overall procedure of algorithm to answer a two-branch enquiry is shown; but there are trees that are more complicated in pointbase' world. In this section, we show DT flexibility and applicability in these trees so that we can answer these trees with seeking of end steps just once.

### 3.1 JPS WITH MORE THAN TWO BRANCHES

As mentioned in primary definition, DT is a table with three columns that first two columns are steps of each branch and its third column is common step between two branches; but in the world, it is possible that several branches were linked together in one JP. For example, assume *Q2: //A[./C][./D]/B;*
Here it is enough that we change primary definition of DT as follows:
*Secondary Definition of DT:* DT is a table with $M+1$ columns for a JP with $M$ sub-branch so that its $1^{st}$ to $M^{th}$ columns are end nodes of branches and last column is common step of JP between all steps.

### 3.2 TREES WITH SEVERAL DIFFERENTS JPS

In a enquiry, each DT will be used for one JP. Therefore, for trees with *M* JPs we need *M* DTs; but these DTs cannot be used independently and there is relationship between them. Therefore, we need two changes: first, we use *DT_Schema* instead of DT.
*Definition:* DT_Schema shows a set of n DT for a enquiry with n JP along with their relations.
*Example:* suppose that we want to build an *DT_Schema* for enquiry. This EH has three branches *(author1,author2,author3)*. The first link point is *B* which links two first branches *A//B/C* and *A//B/D*. *A* is another JP between two first and third branch. Therefore, output of DT will be used as a field. (see figure5)
Second change must be in sequence of steps seeking to generation of Final Data. This change illustrated in figure. This means that at first it seek those JPs that are in lower position in EH tree. The procedure is as follows: when there are orders of seek between several JP, it begins with first JP. A recursive procedure called *Match_Proc* is used that consider orders of seek. If matching seek was successful for a DT. This procedure tests next DT. This seek continues while matching seek is successful for all DT. If matching seek was not successful for one DT, we must do jump from either that or previous DT and matching seek begin from previous DT.

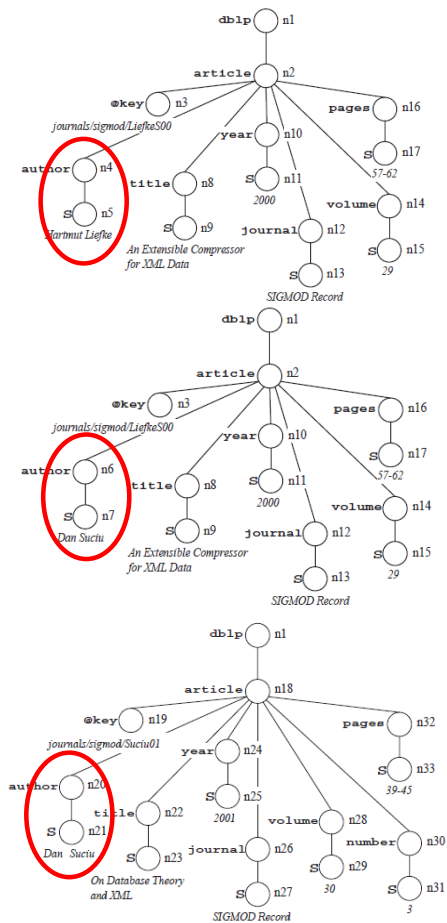

Figure 5. Three type of match process

### 3.3 TREES WITH *, ?, // AND /

DT algorithm is similar to both Query Index and Containment link algorithms. For single-branch trees, query indicators undertake the responsibility of enquiry conversion to absolute query. Fortunately, some of them such as *YAPI*[19] have acceptable performance on various operators (*, ?, // and /) in single-branch trees and don't need to access to real point of the dataset and just with access to PG can answer to various kind of single-branch trees.

## 4. EXPERIMENTAL DATA

In this section we present the Data of our experiments. As discussed above, we categorize the existing pattern enquiry seeking algorithm into three parts. We tested our DT algorithms with *Apriori and Sax*. *Apriori* is selected as the representative of holistic branch link algorithms of Part *B* and Sax as the representative of Part *C*, the algorithms which only access end steps of EH in the Pattern dataset. As mentioned above our DT algorithms are classified into the Part *C* too.

*Our query index:* In second step our algorithm needs to one of query indicators to convert single-branch trees to absolute query of Data steps in PG. There are many query index algorithm to choose; but each algorithm tries to answer to complicated trees by itself. Therefore, for many of trees they need to access to real point of the dataset and thus they have not enough performance whereas in our algorithm a query index is used just on PG and to answer to single-branch trees. Therefore, it must have only two below properties:

   1. Its PG is small and it answers to single-branch trees quickly.

   2. It is applicable for all single-branch trees with all possible operators (*, ?, //)

Among all query index algorithms, the best option that provides two above properties is *YAPI* [19]. It is quickest and cheapest algorithm to answer to single-branch trees.

Pointsets: We use four pointsets *TreeBank[15], XMark[17]* and *DBLP[11]* and a Unknown pointset in our experiments. *DBLP* is a famous pointset which is a shallow and wide dataset. Against DBLP, we use well-known TreeBank pointset which is a deep dataset.

*Unknown pointset*: We build unknown pointset with the depth of *12* and width of step – maximum point of children of a step – *10*. The points tags of this pointset are only *A, B, C, D, E* and *F*. In this way, one point could have one or some homonymous steps as children. As a Data, the path Guide of the dataset could be complex and nested. Here, the numbers, types and orders of children of steps are chosen accidentally.

*Original Hierarchical*: In our experiments, the extended Hierarchical labels are not stored by the dotted-decimal strings displayed (e.g.\*1.2.3.4"*), but rather a compressed binary representation. In particular, we used *UTF-8* encoding as an efficient way to present the integer value, which was proposed by Tatarinov et al. [8].

Trees: In order to test our DT algorithm with Sax, we use trees that are listed in the Table1. Each enquiry has its distinguished property. The enquiry *XQ1* is a single enquiry with *P-C* relationships. For this kind of trees we do not need to generate DT. The trees *XQ4* and *XQ5* are several-branch trees with A-D relationships. The enquiry *XQ3* is also a several-branch enquiry but with P-C relationships and *XQ2* is combination of A-D and P-C relationships.

We choose three parameters to test our DT algorithm with Sax : *i) point of points read, ii) Size of disk files scanned* and *iii) execution time*.

*Point of points read*: In both algorithms, just Ends Nodes of EH will be seeked; but there are two fundamental differences: 1) in Sax at first, each step will be checked whether it has single-branch condition or not; but in our algorithm, we only access those steps, which are member of one enquiry branch. 2) Sax try to answer the enquiry by direct testing of each branch end nodes in dataset and it testes many ends nodes that have not any path relation with each other; but in our algorithm with considering DT, only those end nodes will be tested that have path relation with each other and many steps don't need to be accessed because they have no counterpart in other branch. This difference is more obvious in parent and child trees.

*Size of disk files scanned:* In Sax method when we do test, we need to save some steps because it is possible that they can produce part of answer in test with another step in the future. This is because Sax try to answer the enquiry by direct testing of steps blindly; but in our algorithm, we do not need to save any intermediate point because the way of step seeking and answering the enquiry are specified in DT.

*Execution time*: the execution time of Sax seems to be more than DT. Sax needs to decode the labels to their querys and then test them but in our algorithm, there is no need to decode step labels. Figure 6 confirms the discussion. Our experiments run on a PC with 2.2 GHz Intel Pentium IV seek running Red Hat Linux 8.0 with 2 GB of main memory.

Table 1. Queries used to compare DT with Sax

| Query Name | Query | DataBase |
|---|---|---|
| XQ1 | /site/people/person/gender | XMARK |
| XQ2 | /S[.//VP/IN]//NP | TreeBank |
| XQ3 | /S/VP/PP[IN]/NP/VBN | TreeBank |
| XQ4 | //article[.//sup]//title//sub | DBLP |
| XQ5 | //inproceedings//title[.//i]//sup | DBLP |

Table 2. Queries used to compare DT with Apriori

| Query Name | Query | Dataset |
|---|---|---|
| XQ1 | //dblp/artcle[author]/[.//title]//year | DBLP |
| XQ2 | //people//person[.//address/zipcode]/profile/education | XMark |
| XQ3 | //S//VP/PP[IN]/NP/VBN | TreeBank |

*Apriori* :In this section, we test our algorithm with Apriori algorithm as representative of *B* part algorithm. We test our algorithm with Apriori in two criteria of *i) point of points read* and *ii) execution time*. Trees are in table2, Apriori link all of algorithm in its part will access to all EH steps to answer the enquiry. Therefore, it will have more step access than Sax method to answer the enquiry; but it does not need to convert Hierarchical numbers to query point's name, as a Data, in some cases it operates better than Sax in execution time factor. Figure 7 confirms the discussion.

*Unknown Pointset:* Here we execute our trees on unknown Pointset that is described before. This pointset has many namesake points and a non-uniform schema. Therefore, it shows efficiency of algorithms clear.

*Single-branch trees:* Both *DT* and *Apriori*, execute *8* single-branch trees *A1, A2, ..., A8* with *2, 3 , ..., 9* length respectively. All trees are Partial, i.e, they begin with //, As shown in figure 8, as many as point of single-branch trees steps increase, point of points to be accessed in the dataset in DT decrease.

*Several-branch trees*: Both *DT and Sax*, execute *A1, A2, A3* and *A4* trees which have *2, 3, 4, 5* branches respectively. As shown in figure 8 in both algorithms when number of branches increases, point of step accesses will increase whereas growth rate of DT is very less than growth rate of Sax.

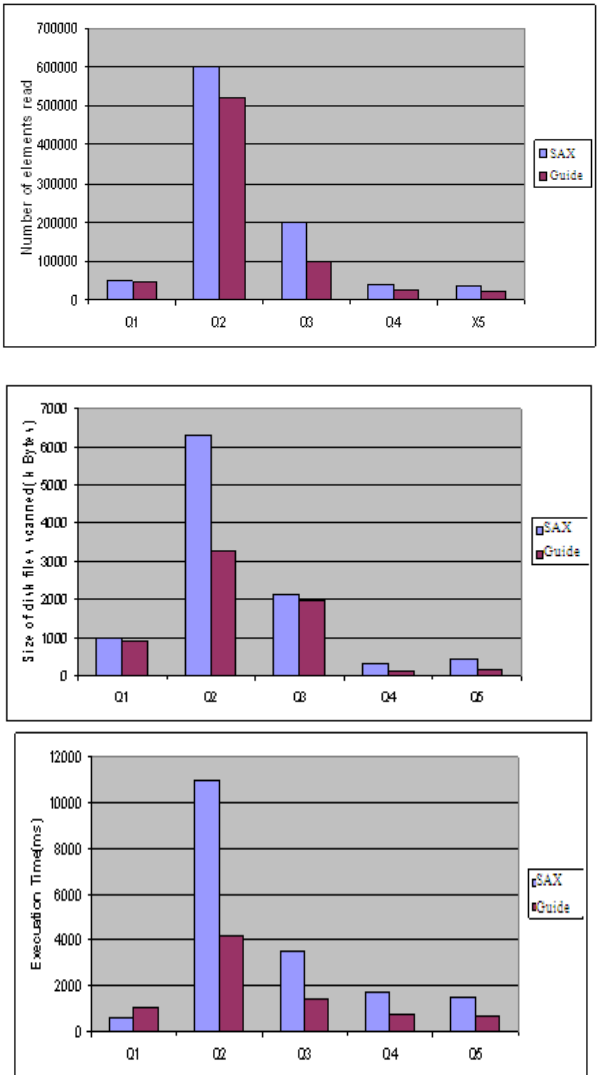

Figure 6. DT(Guide) in Comparison with Sax

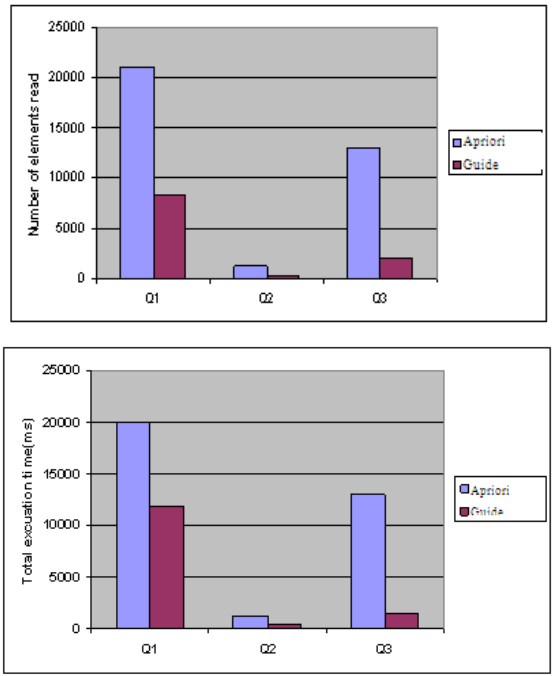

Figure 7. DT (Guide) in Comparison with Apriori

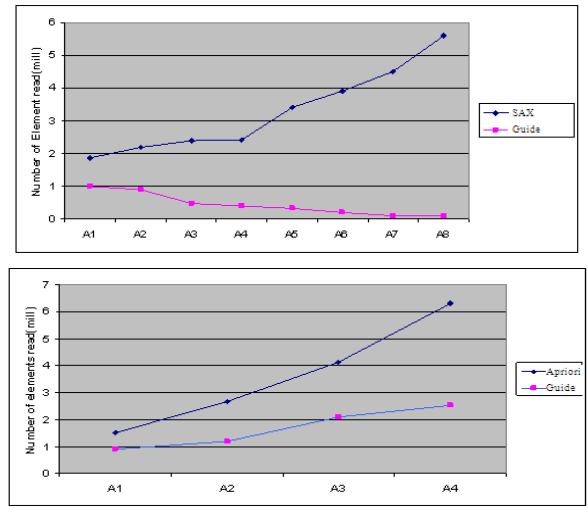

Figure 8. DT and Unknown Pointset

## REFERENCES

[1]. S.Abiteboul, R,Hull, V.vianu, Foundations of Databases. Addison Wesley, 1995.
[2]. Richard Bird, Introduection to Functional Programming using Haskell. Prentice Hall, 1998.
[3]. P.Buneman, M.Fernandez, D.Suciu. UnQl: A Query language and algebra for semistructured data based on structural recursion. VLDB Journal, to appear.
[4]. Catriel Beeri and Yoram Komatzky. Algebraic Optimization of Object-oriented Query Languages.

Theoretical Computer Science 116(1&2):59{94, August 1993.

[5]. Francois Bancihon, Paris Kanellakis, CDelobel. Building an Object-Oriented Database System. Morgan Kaufmann, 1990.

[6]. Peter Buneman,Leonid, Dan Suciu, Vam Tannen, and Limsoon Wong. Comprehension Syntax. SIGMOD Record, 23: 87{96, 1994.

[7]. David Beech, Ashok Malhotra, Michael Rys. A Formal Data Schema and Algebra For XML. W3C XML Query working group note, Septenber 1999.

[8]. Peter Buneman Shamim Naqvi, Val Tannen, Limsoon Wong. Principles of programming with complex abject and collection types. Theoretical Computer Scince 1995.

[9]. Catriel Beeri and Tariv Tzaban, SAL :An Algebra for Semistructures Data and XML. International Workshop on the Web and Databases(WebDB'99). Philadelphia, Pennsylvania, June 1999.

[10]. R.G.Cattell, The Object Database Standard: ODMG 2.0. Morgan Kaufmann, 1997.

[11]. Don Chamberlin, Jonathan Robie, and Daniela Florescu. Quilt: An XML Query Languages for Heterogeneous Data Sources. Intemational Workshop on the Web and Databases(WebDB'2000), Dallas, Texas, May 2000.

[12]. Vassilis Christophides and Sophie Clluet and J_er^ome Sim_eon. On Wrapping Query languages and E_cient XML Integration. Proceedings of ACM SIGMOD Conference on Management of Data, Dallas, Texas, May 2000.

[13]. S.Cluet and G.Moerkotte, Nested queries in object bases,Workshop on Database Programming Languages Pages 226{242,New York,August 1993.

[14]. S.Cluet, S.Jacqmin and J.Sim_eon The New YATL: Design and Speci_cations. Technical Technical Report, INRIA, 1999.

[15]. L.S.Colby, A recursive algebra for nested relations. Information Systems 15(5):567{582, 1990.

[16]. Hugh Darwen(Contributor ) and Chris Data. Guide to the SQL Standard: A User's Guide to the Standard Database Language SQL Addison-Wesley, 1997.

[17]. A.Deutsch.M.Fernandez, D.Florescu. A.Levy, and D.Sueiu. A query language for xml. In International World Wide Web Conference, 1999 , http://www.research.att.com/-mff/files/final.html

[18]. J.A.Goguen, J.W.Thatcher, E.G.Wagner, An initial algebra approach to the speci_cation, correctness, and implementation of abstract data types. In Current Trends in Programming Methodology, pages 80{149, 1978.

[19]. Haruio Hosoya, Benjamin Picrce, XDuce: A Typed XML Prossing language (Preliminary Report) WebDB Workshop 2000.

[20]. M.Kifer, W.Kim, and Y.Sagiv, Querying object-oriented databases. Proccedings of ACM SIGMOD Conference on Management of Data, pages 393{402, San Diego, California, June 1992.

[21]. Leonid Libkin and Limsoon Wong, Query languages for bags and aggregate functions. Journal of Computer and Systems Sciences, 55(2):241{272, October 1997.

[22]. Leonid Libkin, Rona Machlim , and Limsoon Wong, A Query language for multidimensional arrays: Design implementation, and optimization techniques. SIGMOD 1996.

[23]. Al-Khalifa.S., Jagadish. H.V., Koudas.N., Patel. J.M., Srivastava. D., Wu. Y. Structural Joins: A Primitive for Efficient XML Query Pattern Matching. In Proc. ICDE: 141-152(2008)

[24]. Chien. Et. Efficient Structural Joins on Indexed XML, In Proc. VLDB Conference(2009)

[25]. Jiang, H., Wang, W., Lu, H., and Xu Yu, J, an Chin. B, XR-Tree: Indexing XML Data for Efficient Structural Joins. In Proc. ICDE  Conference :253—264(2008).

[26]. Mathis. C., Härder. T, Haustein. M. Locking-Aware Structural Join Operators for XML    Query Processing In Proc SIGMOD Conference: 467 - 478 (2008).

[27]. Jiaheng Lu , Xiaofeng Meng, Tok Wang Ling . Indexing and querying XML using extended Dewey labeling scheme.DATAK-01284;No of pages25(2010).

[28]. Jun Liu.OTwig :An Optimised Twig Pattern Matching Approach for XML Databases. (2010).

[29]. Wu. Y., Patel. M. J., and Jagadish. H. V. Structural join order selection for XML query optimization, In Proc. VLDB Conference,(2008).

[30]. Bruno. N., Koudas. N, Srivastava.D. Holistic Twig Joins: Optimal XML Pattern Matching, In Proc. SIGMOD Conference: 310–321(2007).

[31]. Jiang, H., Wang, W., Lu, H., and Xu Yu, J. Holistic Twig Joins on Indexed XML, In Proc. VLDB Conference :273-284  (2009).

[32]. Sayyed Kamyar Izadi, Mostafa S. Haghjoo. Evaluation of tree-pattern XML queries supported by structural summaries. DATAK 1139 No. of Pages 20, Model 3G. (2008).